\documentclass[conference]{IEEEtran}
\IEEEoverridecommandlockouts

\usepackage{cite}
\usepackage{amsmath,amssymb,amsfonts}
\usepackage{algorithmic}
\usepackage{graphicx}
\usepackage{textcomp}
\usepackage{setspace,url}
\usepackage{numprint}
\usepackage{colortbl,dcolumn}
\npthousandsep{,}
\usepackage[dvipsnames]{xcolor}
\usepackage{makecell}
\usepackage{multirow}
\usepackage{cite}
\usepackage{hyperref}
\usepackage{enumitem}

\def\ninept{\def\baselinestretch{1.144}\let\normalsize\small\normalsize}

\begin{document}
\ninept

\title{The First VoicePrivacy Attacker Challenge}

\author{Natalia Tomashenko$^1$$^*$, Xiaoxiao Miao$^2$, Emmanuel Vincent$^1$$^*$, Junichi Yamagishi$^3$
\thanks{
$^*$This work was supported by the French National Research Agency under project Speech Privacy and project IPoP of the Cybersecurity PEPR.}
\\$^1$\textit{Universit\'e de Lorraine, CNRS, Inria, Loria, F-54000} Nancy, France;
\\$^2$\textit{Singapore Institute of Technology}, Singapore; 
  $^3$\textit{National Institute of Informatics}, Tokyo, Japan 
 \\ natalia.tomashenko@inria.fr, xiaoxiao.miao@singaporetech.edu.sg, emmanuel.vincent@inria.fr,  jyamagis@nii.ac.jp
}

\maketitle

\begin{abstract}

The First VoicePrivacy Attacker Challenge is an ICASSP 2025 SP Grand Challenge which 
focuses on evaluating \textbf{attacker systems against a set of voice anonymization systems} submitted to the VoicePrivacy 2024 Challenge.
Training, development, and evaluation datasets were provided along with a baseline attacker. 
Participants developed their attacker systems in the form of automatic speaker verification systems and submitted their scores on the development and evaluation data. 
The best attacker systems reduced the equal error rate (EER) by 25--44\% relative w.r.t.\ the baseline. 
\end{abstract}


\vspace{-4pt}

\section{Context} 
\label{sec:context}
\vspace{-5pt}

Speech conveys a lot of personal data, e.g., age and gender, health, geographical or ethnic origin, and socio-economic status.
Formed in 2020, the VoicePrivacy initiative 
 \cite{Tomashenko2021CSl} 
promotes privacy enhancing solutions for speech technology via a series of
benchmarking challenges. Privacy preservation is formulated as a game between \emph{users} who process their utterances (referred to as \emph{trial} utterances) with a privacy enhancing system prior to sharing with others, and \emph{attackers} who access these processed utterances and wish to infer information about the users. The level of privacy offered by a given solution is measured as the lowest error rate among all attackers.

The first three VoicePrivacy Challenge editions
\cite{Tomashenko2021CSl,tomashenko2024voiceprivacy} 
focused on improving voice anonymization systems. In particular, the systems submitted to the VoicePrivacy 
2024 Challenge had to: 
    (a) output a speech waveform; 
    (b) conceal speaker identity at the \emph{utterance level}; 
    (c) not distort linguistic and emotional content.
The processed utterances sound as if they were uttered by another \emph{pseudo-speaker}, which is selected independently for every utterance and can be an artificial voice not matching any real speaker.
A \emph{semi-informed attack model} 
was assumed, whereby attackers have access to the voice anonymization system
and seek to re-identify the original speaker behind each anonymized trial utterance. Specifically, an ECAPA-TDNN
automatic speaker verification (ASV) system was 
trained by the participants on data anonymized using their anonymization system. While this attack model is undeniably the most realistic to date, the provided attacker system is not its strongest possible implementation as it does not exploit spoken content similarities, specific pseudo-speaker selection strategies,
or stronger ASV architectures, 
among others.
To ensure a fair and reliable privacy assessment, it is essential to find the strongest possible attacker against every anonymization system. Hence, the current challenge edition takes the attacker's perspective and focuses on the development of attacker systems against voice anonymization systems \cite{tomashenko2024first}.

 \vspace{-4pt}

\section{Task}
\label{sec:task}
\vspace{-4pt}

Participants were required to develop one or more attacker systems against one or more voice anonymization systems selected among three VoicePrivacy 2024 Challenge baselines \cite{tomashenko2024voiceprivacy} and four
systems developed by the VoicePrivacy 2024 Challenge participants. 
For each speaker of interest, the attacker is assumed to have access to one or more utterances spoken by that speaker, which 
are referred to as \textit{enrollment} utterances.
The attacker system shall 
output an ASV score for every given pair of trial utterance and enrollment speaker, where higher (resp., lower) scores correspond to same-speaker (resp., different-speaker) pairs.

To develop and evaluate their attacker system against a given voice anonymization system, in line with the assumed semi-informed attack model, participants had access to:
(1) anonymized trial utterances;
(2) original and
anonymized 
enrollment utterances;
(3)
original and
anonymized training data (as well as other publicly available training resources specified in Section~\ref{sec:data}) for the ASV system;
(4) a 
description of the voice anonymization system;
(5) the software implementation of that 
system when available.

 \vspace{-4pt}

\section{Data}\label{sec:data}
 \vspace{-4pt}

The datasets are presented in Table~\ref{tab:data}.

\vspace{-8pt}

\begin{table}[!th]
\centering
  \caption{Number of speakers and utterances in the attacker training, development, and evaluation sets.}\label{tab:data}
\vspace{-3pt}
 \resizebox{0.48\textwidth}{!}{
  \centering
  \begin{tabular}{|c|l|l|r|r|r|r|}
\hline
 \multicolumn{3}{|c|}{\textbf{Subset}} &  \textbf{Female} & \textbf{Male} & \textbf{Total} & \textbf{\#Utter.}  \\ \hline \hline
 \multirow{1}{*}{Train} & \multicolumn{2}{l|}{ LibriSpeech: train-clean-360} & \numprint{439} & \numprint{482} & \numprint{921} & \numprint{104014} \\ \hline\hline
\multirow{2}{*}{Dev} & LibriSpeech & Enrollment & 15 & 14 & 29 & 343\\ \cline{3-7}
& dev-clean & Trial & 20 & 20 & 40 & \numprint{1978}\\ \cline{1-7}
\multirow{2}{*}{Eval} & LibriSpeech & Enrollment & 16 & 13 & 29 & 438\\ \cline{3-7}
& test-clean & Trial & 20 & 20 & 40 & \numprint{1496}\\ \cline{1-7}
\end{tabular}}
\end{table}
\normalsize

\vspace{-7pt}

\textbf{Training resources.} 
The training set is the \textit{train-clean-360} subset of \textit{{LibriSpeech}}.
In addition, participants were allowed to propose other training resources such as speech corpora and pretrained models before the deadline.
Based on these suggestions, the final list of training resources was published in the evaluation plan~\cite{tomashenko2024first}.

\textbf{Development and evaluation data.}
 The development and evaluation sets comprise \textit{LibriSpeech dev-clean} and \textit{test-clean}.

\textbf{Voice anonymization systems.}\label{sec:anon_systems}
The voice anonymization systems to be attacked include three baseline systems (\textbf{B3}, \textbf{B4}, and \textbf{B5}) \cite{tomashenko2024voiceprivacy,tomashenko2024first} and four selected 
systems developed by the VoicePrivacy 2024 Challenge participants (\textbf{T8-5}, \textbf{T10-2}, \textbf{T12-5}, and \textbf{T25-1}):
 \footnotesize{
\begin{itemize}[leftmargin=*,noitemsep]
    \item \textbf{B3}  --  based on phonetic transcription, pitch and energy modification, and artificial pseudo-speaker embedding generation.

  \item \textbf{B4}  --  based on neural audio codec language modeling. 

 \item \textbf{B5} --  based on vector quantized bottleneck (VQ-BN) features extracted from an ASR model and on original pitch. 
  \item \textbf{T8-5}  \cite{xinyuan2024hltcoe}
  --  random selection of one of two methods for each utterance (with probability $p$ for the second method): (1) a cascaded ASR-TTS system with \textit{Whisper} 
  for ASR and \textit{VITS} 
  for TTS  and (2) a k-nearest neighbor (kNN) voice conversion (VC) system operating on \textit{WavLM} 
  features.

\item \textbf{T10-2}  \cite{yao2024npu}
--  neural audio codec, with specific disentanglement of 
linguistic content, speaker identity and emotional state.

\item \textbf{T12-5} \cite{Kuzmin2024ntu}
-- based on \textbf{B5}, with
additional pitch smoothing.
\item \textbf{T25-1} \cite{Gu2024ustc}
-- disentanglement of content (VQ-BN as in \textbf{B5}) and style (global style token (GST) 
features
and emotion transfer
  from target speaker utterances.
\end{itemize}
}
\normalsize
The code of \textbf{B3}, \textbf{B4}, and \textbf{B5} is available 
and could be used to develop attacker systems by, e.g., generating different or additional training data to train those systems.

\section{Evaluation metric}
\label{sec:metric}

We use the equal error rate (EER) metric to evaluate the attacker's performance. This metric has been used in all VoicePrivacy Challenge editions. 
The lower this metric, the stronger the attacker.
The number of same-speaker and different-speaker trials in the development and evaluation datasets is given in Table~\ref{tab:trials}. The attackers were  ranked separately for each voice anonymization system.

 \vspace{-5pt}

\begin{table}[htbp]
  \caption{Number of speaker verification trials.}\label{tab:trials}
\vspace{-3pt}
  \centering
   \resizebox{0.48\textwidth}{!}{
  \begin{tabular}{|l|l|l|r|r|r|}
\hline
 \multicolumn{2}{|c|}{\textbf{Subset}} & \textbf{Trials} &  \textbf{Female} & \textbf{Male} & \textbf{Total}  \\ \hline \hline
\multirow{2}{*}{Dev} & LibriSpeech & Same-speaker & 704 & 644 & \numprint{1348} \\ \cline{3-6}
 & dev-clean & Different-speaker	& \numprint{14566} & \numprint{12796} &	\numprint{27362} \\ \cline{1-6}
\multirow{2}{*}{Eval} & LibriSpeech & Same-speaker & 548 & 449	& \numprint{997} \\ \cline{3-6}
  & test-clean & Different-speaker & \numprint{11196} & \numprint{9457} &	\numprint{20653} \\ \cline{1-6}
  \end{tabular}}
\end{table}

\vspace{-8pt}

\section{Baseline attacker system}\label{sec:baseline_attack}

\begin{figure}[h!]
\centering
\includegraphics[width=85mm]{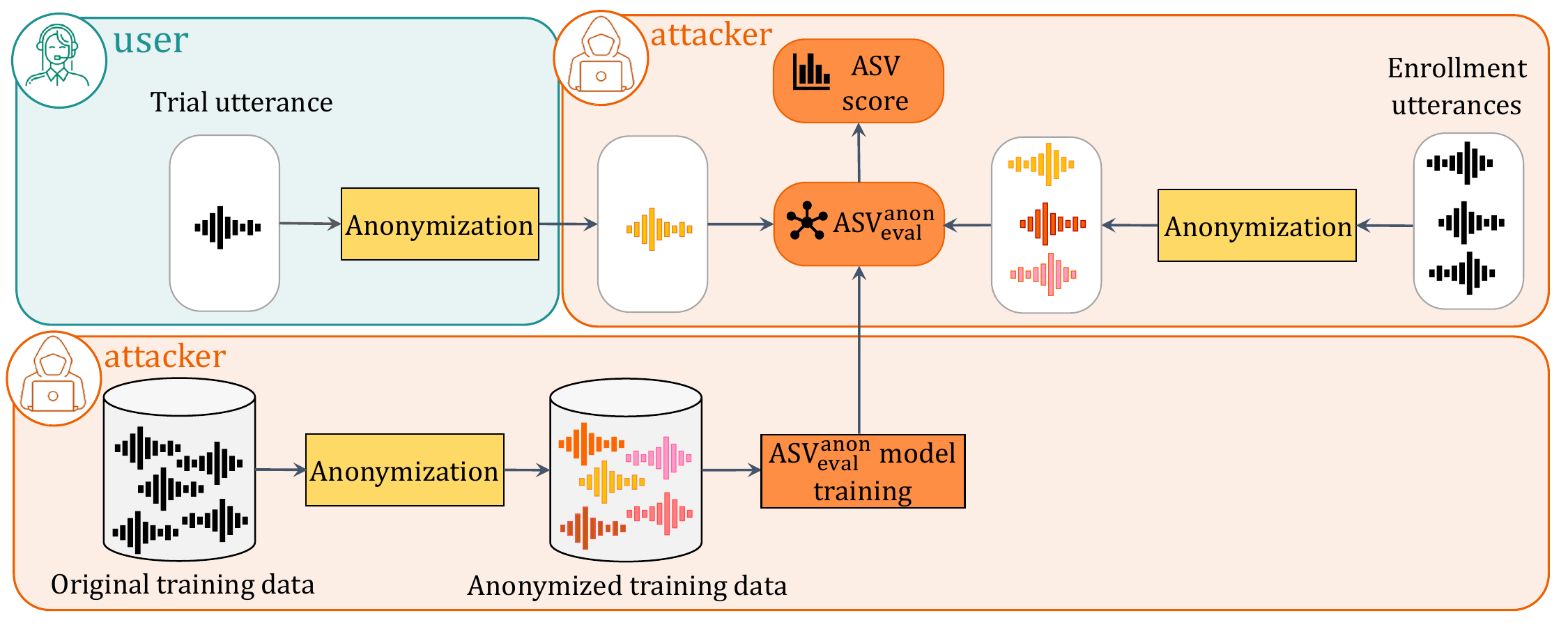}
\caption{Baseline attacker: training $ASV_\text{eval}^{\text{anon}}$ on anonymized training data and using it to compare anonymized trial and enrollment data.}
\vspace{-8pt}
\label{fig:asv-eval}
\end{figure}

As a baseline, we consider the attacker system used in the VoicePrivacy 2024 Challenge \cite{tomashenko2024voiceprivacy} (see Fig.~\ref{fig:asv-eval}).
The ASV system (denoted $ASV_\text{eval}^{\text{anon}}$) is an ECAPA-TDNN 
with 512 channels in the convolution frame layers, implemented by adapting the \textit{SpeechBrain} 
\textit{VoxCeleb} recipe to \textit{LibriSpeech}, and it is trained on  anonymized training data.
For a given trial utterance and enrollment speaker, the attacker computes the average speaker embedding of all anonymized enrollment utterances from that speaker and compares it to the speaker embedding of the anonymized trial utterance.

\vspace{-6pt}

\section{Challenge results and conclusions}

\begin{figure}[h!]
\centering
\includegraphics[width=87mm]{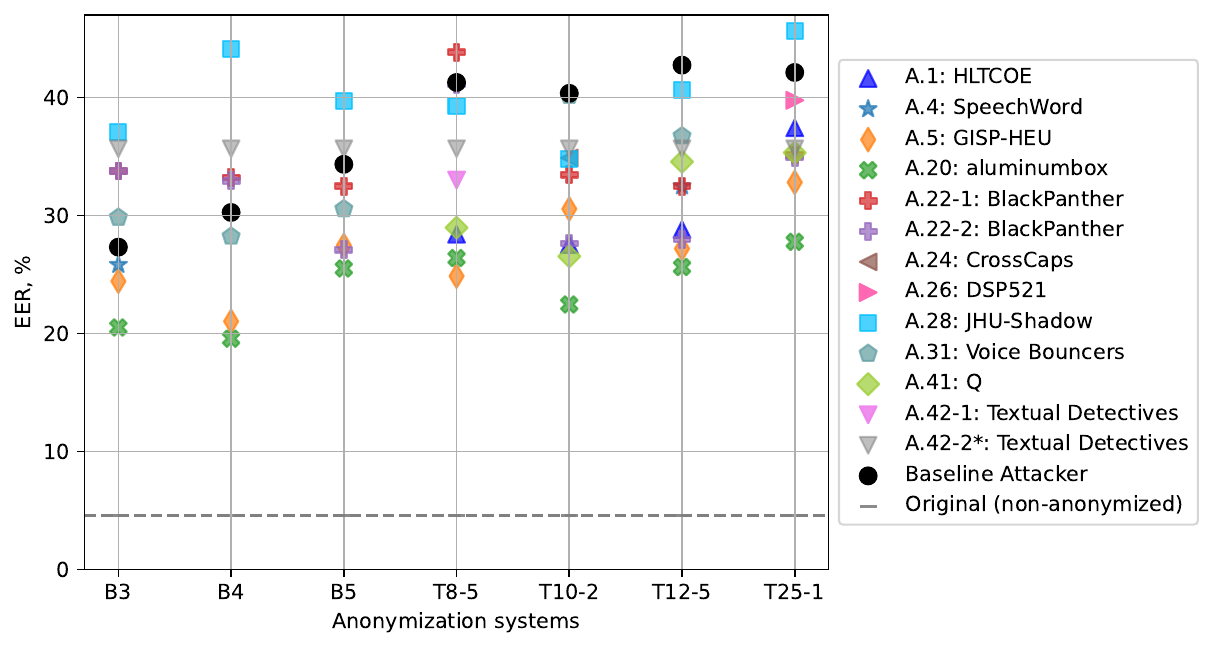}
\vspace{-8pt}
\caption{Challenge results  on the evaluation dataset.}
\label{fig:res}
\end{figure}

The challenge attracted 41 registered teams from academia and industry in 11 countries. Among them, 
11 teams successfully submitted their results (55 submissions for 7 anonymization systems), which are summarized in Fig.~\ref{fig:res}. Many attackers significantly outperform the baseline attacker.
The best ones reduce EER by 7--18\% absolute (25--44\% relative) for different anonymization systems\footnote{The attacker \textbf{A.42-$\textbf{2}^*$} uses only textual data for training. It does not comply with the challenge rules due to the use of an undeclared BERT model.}.

The best attacker was developed by team $\textbf{A.5}$ \cite{team5_gisp-heu} for anonymization system \textbf{T8.5} and by team $\textbf{A.20}$ \cite{team20_aluminumbox} for every other anonymization system. 
 $\textbf{A.20}$ adapts a pretrained \textit{ResNet34} ASV model from \href{https://github.com/wenet-e2e/wespeaker}{\textit{WeSpeaker}} using the \textit{LoRA} technique on the provided anonymized data.
  $\textbf{A.5}$ for \textbf{T8-5} proposes to use a so-called \textit{ECAPA-PLDA-Mix} model, which combines an \textit{ECAPA-TDNN} feature extractor trained on mixed datasets with a \textit{PLDA}-based scoring module trained on anonymized data, and \textit{SpecAugment} data augmentation. 
Other successful attackers' strategies include, among others, using
a proposed \textit{SpecWav} attack based on the \textit{wav2vec2.0} feature extractor and 
 spectrogram resizing (\textbf{A.41}) \cite{team41_q}; 
fine-tuning the \href{https://huggingface.co/nvidia/speakerverification_en_titanet_large}{\textit{TitaNet-Large}} model on anonymized data (\textbf{A.22-2}) \cite{team22_blackpanther};
using alternative distance metrics and voice \textit{kNN-VC}-based voice normalization
(\textbf{A.1}) \cite{team1_hltcoe}. 
The findings of the  Attacker Challenge  reveal that the privacy protection offered by the best anonymization systems from the VoicePrivacy 2024 Challenge was overestimated.
They also show that, while attackers have made great progress in reducing the baseline EERs, the best anonymization methods still provide moderate
protection against speaker re-identification ($\text{EER}>$ 25\%).
A paper with a more detailed analysis of the results will be published in the future.

\vspace{-6pt}

\vspace{-7pt}

\bibliographystyle{IEEEbib}
\vspace{-5pt}

\bibliography{refs-camera-ready}

\end{document}